\documentclass[preprint,aps]{revtex4}

\usepackage{graphicx}
\usepackage{dcolumn}
\usepackage{bm}

\begin{document}

\begin{titlepage}

\title{Edge Stability of BN sheets and Its Application for Designing Hybrid BNC Structures}

\author{Bing Huang$^1$, Hoonkyung Lee$^2$, Bing-Lin Gu$^1$, Feng Liu$^3$, and Wenhui Duan$^1$\footnote{E-mail: dwh@phys.tsinghua.edu.cn}}

\address{$^1$Department of Physics, Tsinghua University, Beijing 100084, P. R. China}
\address{$^2$Department of Physics, University of California, Berkeley, California 94720, USA}
\address{$^3$Department of Materials Science and Engineering, University of Utah, Salt Lake City, Utah 84112, USA}

\date{\today}

\begin{abstract}

First-principles investigations on the edge energies and edge
stresses of single-layer hexagonal boron-nitride (BN) are presented.
The armchair edges of BN nanoribbons (BNNRs) are more stable in energy than
zigzag ones. Armchair BNNRs are under compressive edge
stress while zigzag BNNRs are under tensile edge stress. The
intrinsic spin-polarization and edge saturation play important roles
in modulating the edge stability of BNNRs. The edge energy
difference between BN and graphene could be used to guide the design of the specific hybrid BNC structures: in an armchair BNC nanoribbon (BNCNR), BN domains are expected to grow outside of C domains, while the opposite occurs in a zigzag BNCNR. More importantly, armchair BNCNRs can reproduce unique electronic properties of armchair graphene nanoribbons
(GNRs), which are expected to be robust against edge functionalization or disorder. Within certain C/BN ratio, zigzag BNCNRs may exhibit
intrinsic half-metallicity without any external constraints. These
unexpected electronic properties of BNCNRs may offer unique
opportunities to develop nanoscale electronics and spintronics
beyond individual graphene and BN. Generally, these principles for
designing BNC could be extended to other hybrid nanostructures.

\end{abstract}

\maketitle

 \draft

\vspace{2mm}

\end{titlepage}

\section{Introduction}

Both single-layer boron-nitride (BN) sheet and graphene are
two-dimensional (2D) crystals. Different from graphene, a zero-gap
semimetal, BN sheet displays insulating characteristics due to the large
ionicity of B and N atoms. Few-layers BN was first obtained by
decomposition of borazine on metal surfaces with a matching lattice
or in a mesh structure in the case of a lattice
mismatch\cite{Nagashima-Corso}. Much recent efforts have been made
to synthesize single-layer BN. Free-standing BN single layers were
fabricated in experiments via controlling energetic electron beam
irradiation through a sputter process\cite{Jin-Meyer-2009}. Similar
to the graphene nanoribbons (GNRs), which are patterned from
graphene via lithographical methods\cite{Han-2007, Chen-2007,
Tapaszto-2008}, it is possible to obtain BN nanoribbons (BNNRs) by
cutting single-layer BN sheet.

The unusual electronic and magnetic properties of GNRs and BNNRs
have been widely studied: both theoretical and experimental
investigations reveal that all narrow GNRs are semiconducting
regardless of their chirality (edge shapes)\cite{Castro Neto-2009,
Son-2006, Han-2007, Wang-2008}; meanwhile, zigzag GNRs (ZNGRs) are
predicted to have magnetic ground states, while armchair GNRs
(AGNRs) are nonmagnetic\cite{Son-2006}. Similar to AGNRs, armchair
BNNRs (ABNNRs) also display nonmagnetic semiconducting behavior
independent of their width\cite{Park-2008}; different from ZGNRs,
however, zigzag BNNRs (ZBNNRs) could be either magnetic or
nonmagnetic depending sensitively on their edge
passivation\cite{Park-2008, Zheng-2008, Barone-2008}.

Besides their electronic properties, further understanding of the
edge stability of GNRs and BNNRs is necessary and important for
practical device applications. The edge of a 2D structure is in
analogy to the surface of a 3D structure and the edge stability can
be potentially understood through two fundamental thermodynamic
quantities: edge energy and edge stress, which defines the chemical
and mechanical edge stability, respectively. The two quantities may
interplay with each other affecting various edge-related phenomena
in 2D structures. Some attention has already been paid to the edge
instability of GNRs both in experiments\cite{Meyer-Gass-Liu-Huang} and theories\cite{Shenoy-2008, Bets-2009, Huang-2009}. It is found that AGNRs are more stable than ZGNRs in edge energy, and both are under intrinsic compressive edge stress, which results in edge twisting and warping
instability\cite{Shenoy-2008, Bets-2009, Huang-2009}. Edge
reconstruction or edge saturation could effectively lower the edge
energies as well as relieve the edge compression and hence to
stabilize the edge structures of GNRs\cite{Bets-2009, Huang-2009}.
Comparing with GNRs, BNNRs have similar geometry but quite different
electronic properties, and thus we expect some essential differences
in the edge stability between BNNRs and GNRs. In particular, the
edge stability of BN structures have not been fully explored yet.

In this article, we systematically study the edge energies and edge
stresses of BN sheet using first-principles calculations. Our results show
that ABNNRs are more stable in energy than ZBNNRs. ABNNRs are under
compressive stress, but ZBNNRs are under tensile stress. More
interestingly, the intrinsic spin-polarization and edge adsorption of
H could effectively stabilize the edges of BNNRs. Furthermore, using
the edge energy difference between BN and graphene, we develop
some basic principles for designing the structures of hybrid
BNC\cite{Ci-2009}. Specifically, taking BNC nanoribbons (BNCNRs) as
model systems, we find that in armchair BNCNRs (ABNCNRs), BN
components are expected to form the armchair edges (with C
components inside) to attain the lowest energy; while, C edges are
preferred over BN edges in zigzag BNCNRs (ZBNCNRs). The hybrid BNCNRs
show rich electronic and magnetic properties depending on their
structures and the C/BN ratio. ABNCNRs can reproduce the basic
electronic properties of AGNRs, and ZBNCNRs can exhibit
half-metallicity within certain C/BN ratio. Thus, the hybrid BNC
structures offer more opportunities than individual graphene and
BN for building nano-scale electronics and spintronics.

\section{Computational Methods and Models}

Our calculations were performed using density functional theory
(DFT) in the generalized gradient approximation, with the
Perdew-Burke-Ernzerhof (PBE) functional\cite{PBE} for electron
exchange and correlation potentials, as implemented in the VASP code
\cite{VASP}.  The electron-ion interaction was described by the
projector augmented wave (PAW) method\cite{PAW}, and the energy
cutoff was set to 500 eV.  The structures were fully optimized using
the conjugate gradient algorithm until the residual atomic forces to
be smaller than 10 meV/\AA. The supercell with periodic boundary
conditions was adopted to model the nanoribbons, with a vacuum layer
larger than 15 \AA~ to eliminate the interaction between the ribbon
images in the neighboring cells.

\begin{figure}[tbp]
\includegraphics[width=8.0cm]{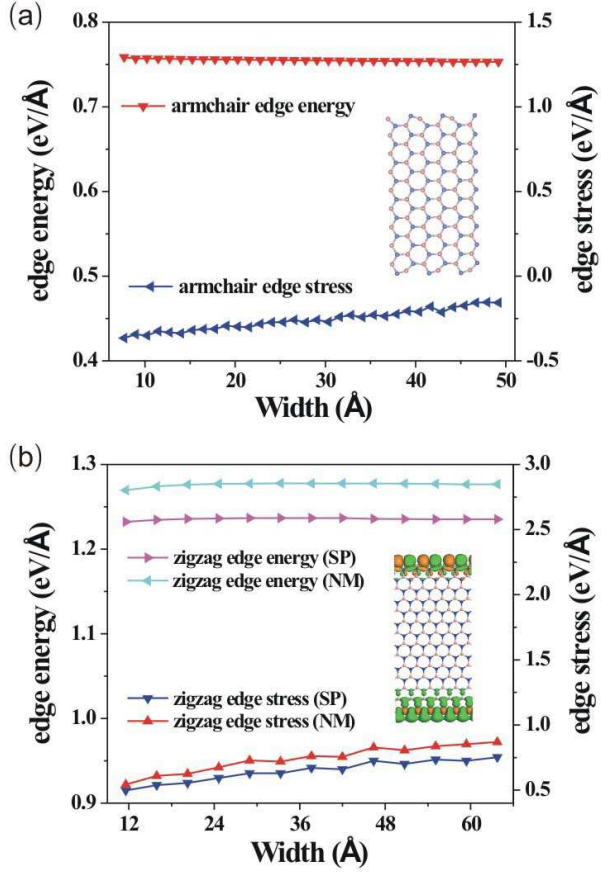}
\caption{(a) The edge energies and edge stresses of ABNNRs as a
function of ribbon width. Inset: schematics of the optimized ABNNR.
Blue and pink balls represent N and B atoms, respectively. (b) The
spin-polarized (SP) and nonmagnetic (NM) edge energies and edge
stresses of ZBNNRs as a function of ribbon width. Inset: schematics
of the optimized ZBNNR associated with the spatial distribution of
spin density of SP state. Green and orange colors represent the
spin-up and spin-down states, respectively.}
\end{figure}

\section{Results and discussion}

\textbf{\emph{Edge Stability of BN Sheets}.} Figure 1a shows the
edge energies and edge stresses of bare ABNNRs with the width
ranging from 7.5 to 50 \AA. Herein, the edge energy of a bare BNNR
is calculated as
\begin{equation}E_{\rm edge}  = \frac{{E_{\rm BNNR}  - \frac{N}{2}\varepsilon _{\rm BN}
}}{{2L}} \end{equation} where $E_{\rm BNNR}$ denotes the total
energy of a BNNR with $N$ boron and $N$ nitrogen atoms in the
supercell, $\varepsilon_{\rm BN}$ is the energy of a pair of BN
atoms in a perfect BN sheet, and $L$ is the length of an edge. The
edge stress is extracted from the calculated stress tensor by
Nielsen-Martin algorithm\cite{Martin-1985}, as described in previous
work\cite{Huang-2009}. The edges of bare ABNNR are reconstructed
(inset structure of Figure 1a): all B atoms at the edge relax inward
and adjacent N atoms are outward away from the edge. The ground
states of bare ABNNRs are nonmagnetic because the dangling-bonds are
passivated after edge reconstruction. The average edge energy of
ABNNRs, $\sim$ 0.75 eV/\AA, displays a weak width dependence, and is
smaller than that of AGNRs ($\sim$ 1.0 eV/\AA)\cite{Huang-2009,
Koskinen-2008}. The edge stresses of ABNNRs are negative (i.e.,
compressive stress) and oscillate weakly ($\sim$ 0.02 eV/\AA) with
increasing ribbon width. The average edge stress of ABNNRs ($\sim
-0.25$ eV/\AA) is much smaller than that of AGNRs ($\sim -1.45$
eV/\AA)\cite{Huang-2009}, indicating that the armchair edges of BN
are mechanically more stable than that of AGNRs. Moreover, the weak
width-dependence of edge energies and edge stresses in ABNNRs are
quite different that of AGNRs\cite{Huang-2009}.

Figure 1b shows the calculated edge energies and edge stresses of
ZBNNRs for different ribbon width (11 $\sim$ 64 \AA). The
nonmagnetic edge energy is independent of ribbon width ($\sim$
1.28 eV/\AA), but the nonmagnetic edge stress increases by $\sim$
0.3 eV/\AA~ as the ribbon width increases from $\sim$ 11 to $\sim$
64 \AA. The positive edge stress means that ZBNNRs are under
intrinsic tensile stress, which is evidently different from that of
ABNNRs. The ground states of bare ZBNNRs are spin-polarized with
antiferromagnetic spin ordering at B-edge (i.e., the outmost atoms
at the edge are B atoms) and ferromagnetic spin ordering at N-edge
(i.e., the outmost atoms at the edge are N atoms), as shown in the
inset of Figure 1b, in agreement with previous
predictions\cite{Zheng-2008, Barone-2008}. The magnetic moment is
$\sim$ 1 $\mu$B per edge B or N atom. The spin-polarized edge
energies are $\sim$ 0.04 eV/\AA~ lower than the nonmagnetic ones.
Furthermore, spin-polarization has a sizable effect of reducing the
stress by $\sim$ 0.1 eV/\AA.

\begin{figure}[tbp]
\includegraphics[width=8.0cm]{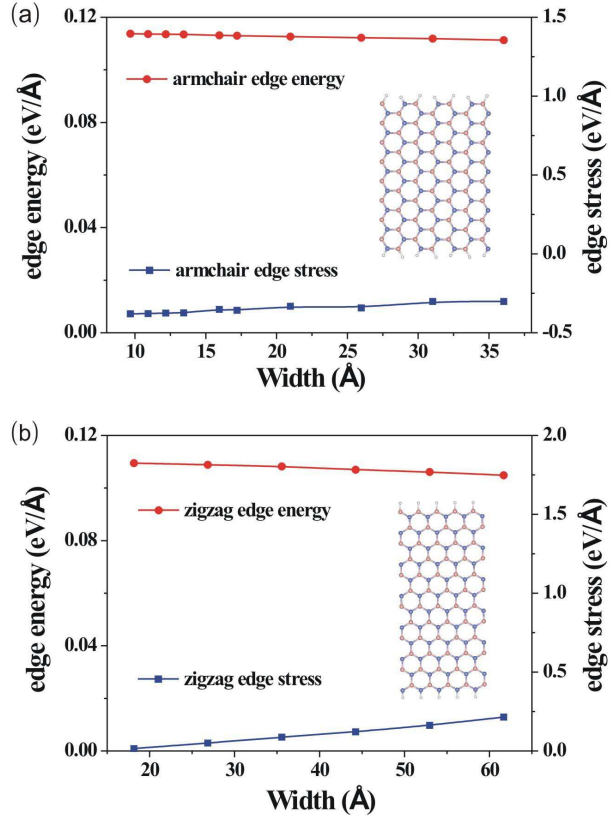}
\caption{The edge energies and edge stresses of (a) H-terminated
ABNNRs and (b) H-terminated ZBNNRs as a function of ribbon width.
Insets: schematics of the optimized ABNNR and ZBNNR.}
\end{figure}

Previous studies demonstrated that both armchair and zigzag edges of
graphene are under intrinsic compressive stress, resulting in edge
twisting and warping instability\cite{Shenoy-2008, Bets-2009,
Huang-2009}. Different from graphene, edge warping is expected to
appear only at the armchair edges of BN (compressive edge stress)
but not at the zigzag edges of BN (tensile edge stress). This
indicates that the armchair and zigzag edges of BN should be
distinguishable by their different edge morphologies (undulations)
in experiments.

It is known that edge adsorption of H in GNRs can relieve the edge
compression and lower the edge energies\cite{Koskinen-2008,
Bets-2009, Huang-2009}, because the compressive edge stress of GNRs
originate from the dangling bonds of bare edge atoms. Thus, we
further investigate the effect of hydrogen passivation on the edge
stability of BNNRs. The edge energy of a H-terminated BNNR is
calculated as
\begin{equation}
E_{\rm edge}  = \frac{{E_{\rm H-BNNR}  - \frac{N}{2}\varepsilon _{\rm BN}  -
n_{\rm H}\mu_{\rm H} }}{{2L}}
\end{equation}
where $E_{\rm H-BNNR}$, $N$, $\varepsilon _{\rm BN}$, and $L$ have
the same definition as in Eq. (1). $\mu_{\rm H}$ is the chemical
potential of hydrogen and $n_{\rm H}$ is the number of H atoms in a
supercell. Here, $\mu _{\rm H}$ is calculated as half of the total
energy of an isolated H$_{2}$ molecule. As shown in Figure 2a, the
average edge energy of ABNNRs decreases largely from $\sim$ 0.75
eV/\AA~ to $\sim$ 0.11 eV/\AA~ after H adsorption, but the edge
stress changes little. For ZBNNRs, the average edge energy is around
0.11 eV/\AA, almost the same as ABNNRs; the edge stresses are
reduced largely by $\sim$ 0.50 eV/\AA, as shown in Figure 2b. Notably,
the edge passivation destroys the magnetic states of ZBNNRs, and the
ground states is now nonmagnetic \cite{Zheng-2008}. The above
results strongly indicate that edge saturation could be used to
stabilize the edges of BNNRs, similar to the case of GNRs.

\textbf{\emph{Geometries, Electronic and Magnetic Properties of
Hybrid BNC Sheets}.} The edge energy difference between BN and
graphene is useful to analyze the stability of hybrid BNC
structures. Under equilibrium growth conditions, BN and C are
thermodynamically immiscible, preferring to separate into domains in
planar BNC structures\cite{Yuge-2009, Ci-2009}. For armchair edges,
since the edge energies of ABNNRs ($\sim$ 0.75 eV/\AA) are much
lower than those of AGNRs ($\sim$ 1.0 eV/\AA), BN prefers to form the
edges while C stays inside to lower the overall energy. However, the
situation is totally different for zigzag edges of BNC structures, as
C prefers to form the edges with BN staying inside, because of the
lower edge energies of ZGNRs. It should be noted that the B, N, and C atoms in BNC sheet containing substitutional C atoms were identified in a
very recent experiment\cite{Krivanek-2010}. Therefore, we highly
expect the future experiments to identify the edge atomic species of
hexagonal BNC structures to support our prediction. In the
following, we will take BNCNRs as examples to verify these simple
predictions in theory and systematically study the electronic
properties of BNCNRs.

\begin{figure}[tbp]
\includegraphics[width=8.0cm]{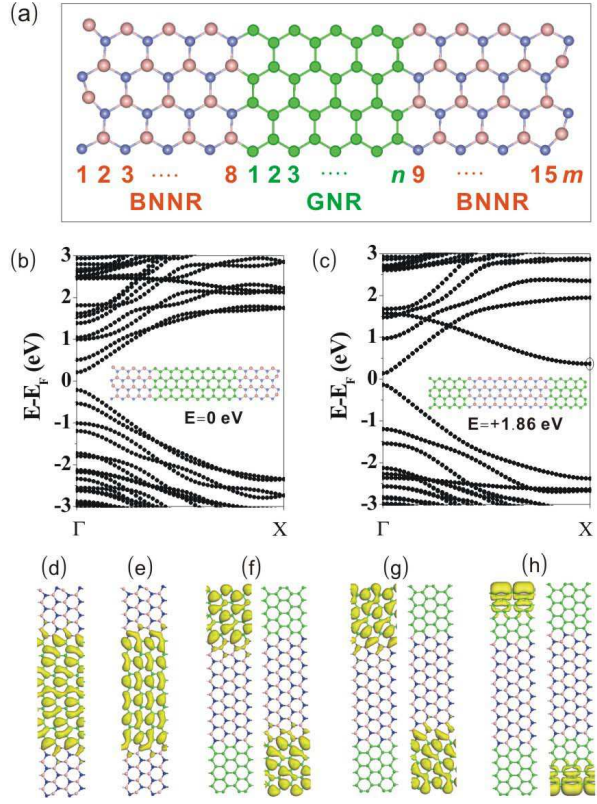}
\caption{(a) The schematic structure of (\emph{m}, \emph{n}) ABNCNR
with \emph{m} and \emph{n} dimer lines for the BNNR and GNR,
respectively; Blue, pink, and green balls represent N, B, and C
atoms, respectively. (b) and (c) are the band structures of (16, 16)
ABNCNR-BN and (16, 16) ABNCNR-CC, respectively. The Fermi level is
set to zero. The corresponding optimized atomic structures and
relative total energies are also shown as an inset in (b) and (c).
(d) and (e) are partial charge densities of the top valence band and
the bottom conduction band at the $\Gamma$ point of (16, 16)
ABNCNR-BN, respectively. (f) and (g) are partial charge densities of
the top valence band and the bottom conduction band at the $\Gamma$
point of (16, 16) ABNCNR-CC, respectively. (h) Partial charge
density of the bottom conduction band at the $X$ point of (16, 16)
ABNCNR-CC.}
\end{figure}

As shown in Figure 3a and Figure 5a, a simple BNCNR consists of BNNR
domains and GNR domains. Following the conventional notations of
\emph{m}-BNNR and \emph{n}-GNR, an armchair (a zigzag) BNCNR is
defined as (\emph{m}, \emph{n}) ABNCNR (ZBNCNR), with \emph{m} and
\emph{n} dimer lines (zigzag chains) across the BN and C ribbon
width, respectively. For example, the structures in Figure 3a and Figure 5a
are referred as armchair (16, 8) ABNCNR an (6, 6) ZBNCNR, respectively.
In this article, we focus on the specific BNCNR configurations that either
BN or C grows at both ribbon edges. In addition, we concentrate our
attention on these situations that BN (or C) stay at both BNCNR
edges symmetrically, since our test calculations indicate that
symmetrical distributions of BN (or C) at both edges are more favorable
in energy than other asymmetrical ones.

In the following, (16, 16) ABNCNRs are taken as examples for BNCNRs with
armchair edges. The component of BNNRs (or GNRs) could be either outside or
inside, namely, the edge atoms could be either C (ABNCNR-CC) or BN
(ABNCNR-BN). The calculated total energies in the two cases indeed
confirm our analysis based on the large edge energy differences
between graphene and BN: ABNCNR-BN is 1.86 eV per unit cell ($\sim$
0.22 eV/\AA) lower in energy than ABNCNR-CC, indicating that
the differences of BN-C domain interfacial energy as well as
inter-domain interaction energy for the two cases are small ($\sim$
0.03 eV/\AA)\cite{Ci-2009}.

The calculated band structures, as shown in Figure 3b and c,
demonstrate that both structures are nonmagnetic semiconductors with
a direct bandgap of $\sim$ 0.42 eV (Figure 3b) and $\sim$ 0.27 eV
(Figure 3c). BNNRs are insulating with a bandgap $>$ 4 eV (based on
our calculation), which is much larger than that of
GNRs\cite{Son-2006, Qimin-2007}, so BN domain in a BNCNR may act as
``energy barrier'' to confine the states of GNR component around the
Fermi level. Similar physical mechanism has been found in partially
hydrogenated graphene\cite{Singh-Xiang-2009}, but precise control of
the distribution of hydrogen atoms on the graphene surface is rather
challenging. The partial charge density analysis confirms that the
bands around the Fermi level are mostly localized in the GNR
domains, as shown in Figure 3d-3g [the states of (16, 16) ABNCNR-CC
around the Fermi level are doubly-degenerate]. Moreover, a state
near the bottom of conduction band in Figure 3c has a small dispersion
with a flattened tail near the X-point and the partial charge
density of the two doubly-degenerate states at X-point are localized
around the two armchair edges of GNR domains (Figure 3h), belonging to
the ``edge states'', which may be caused by the chemical potential
difference between the two boundaries of GNR domain. The edge states
were only found at the zigzag edges of graphene previously\cite{Nakada-1996}, and have exhibited plenty of important
applications\cite{Castro Neto-2009, Son-Kan, Woo-Palacios}.
Surprisingly, our results here demonstrate that it is also possible
to introduce edge states at the armchair edges of graphene via
hybrid BNC structures. Furthermore, doping electrons to the system
(or other chemical functionalization) may pull down the armchair
edge states to the Fermi level for further applications. Although
the structures of ABNCNRs-CC are less stable in energy than
ABNCNRs-BN, it is still possible to realize them in experiments
under non-equilibrium growth conditions via controlling the
deposition rate, temperature and the effect of
substrate\cite{Ci-2009}.

Since the states of ABNCNRs around the Fermi level mainly originate
from GNR domains, we expect that the well known features of AGNRs
could be reproduced in ABNCNRs. For instance, it is known that the
energy gaps of AGNRs decrease as ribbon width increases and exhibit
three distinct family behaviors\cite{Son-2006, Qimin-2007}, and we
expect the similar quantum confinement effect could also exist in
ABNCNRs. The calculated band gaps of (16, \emph{n}) ABNCNRs-BN with
different \emph{n} (width of the GNR part) are shown in Figure 4a.
The band gaps of hybrid ABNCNRs could be divided into three groups
and decrease with the increase of the width, similar to the case of
AGNRs. Different from the gap size hierarchy of AGNRs
($\bigtriangleup_{3p+1}$ $>$ $\bigtriangleup_{3p}$ $>$
$\bigtriangleup_{3p+2}$)\cite{Son-2006, Qimin-2007}, the hierarchy
of ABNCNRs is $\bigtriangleup_{3p}$ $>$ $\bigtriangleup_{3p+1}$ $>$
$\bigtriangleup_{3p+2}$. The discrepancy may come from the essential
differences between the edges of GNRs and the boundaries of BNCNRs,
as found in previous work\cite{Du-2009, Bing-2010}. Besides, the
bandgap values of ABNCNRs are much smaller than those of the
corresponding AGNRs. The band structures of (16, \emph{n})
ABNCNRs-BN ($n= 6$, 7, and 8) are shown in Figure 4b as examples.

\begin{figure}[tbp]
\includegraphics[width=8.0cm]{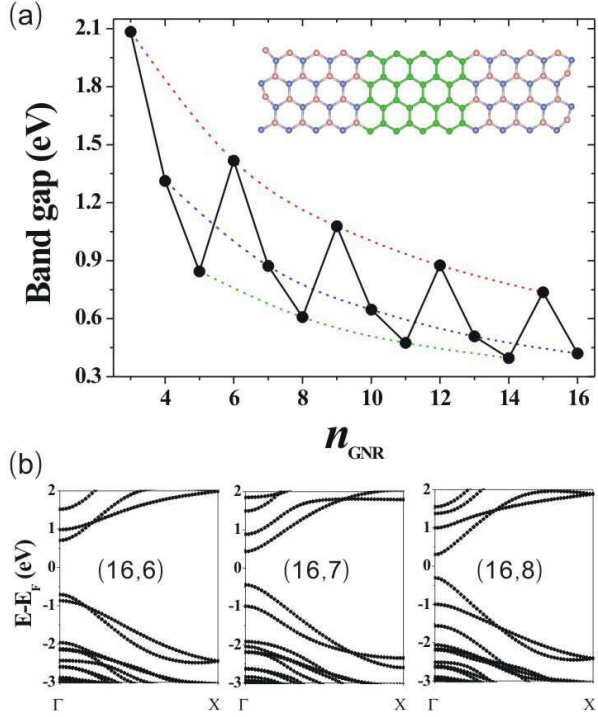}
\caption{(a) The band gaps of (16, \emph{n}) ABNCNRs-BN as a function
of \emph{n} (the width of the GNR part). (b) Band structures of (16,
\emph{n}) ABNCNRs-BN (\emph{n} $=$ 6, 7, and 8). The Fermi level is set to zero.}
\end{figure}

Theoretical studies have predicted interesting electronics of GNRs
with perfect edges\cite{Castro Neto-2009}. Unfortunately, most of
the intrinsic properties predicted have not been observed in
experiments until now, because it is very difficult to get smooth
edges under current experimental technology\cite{Han-2007,
Chen-2007}. The rough edge structures will dramatically influence
the electronic properties of GNRs\cite{Bing-2008-2007-JPCC, Castro
Neto-2009}. Besides, edge chemical functionalization as well as
doping is also inevitable in experiments due to the high chemical
reactivity of the graphene edges\cite{Han-2007, Chen-2007, Li-2008}.
Quite encouragingly, our study strongly implies that ABNCNRs can
exhibit novel electronic properties of prefect GNRs: since the BN
prefers to grow outside of GNRs and the electronic properties of BN
are robust against different chemical
functionalization\cite{Chen-2004,Wu-2006}, the electronic properties
of ABNCNRs, mimicing those of AGNRs around the Fermi level, are
expected to be also robust.

\begin{figure}[tbp]
\includegraphics[width=8.0cm]{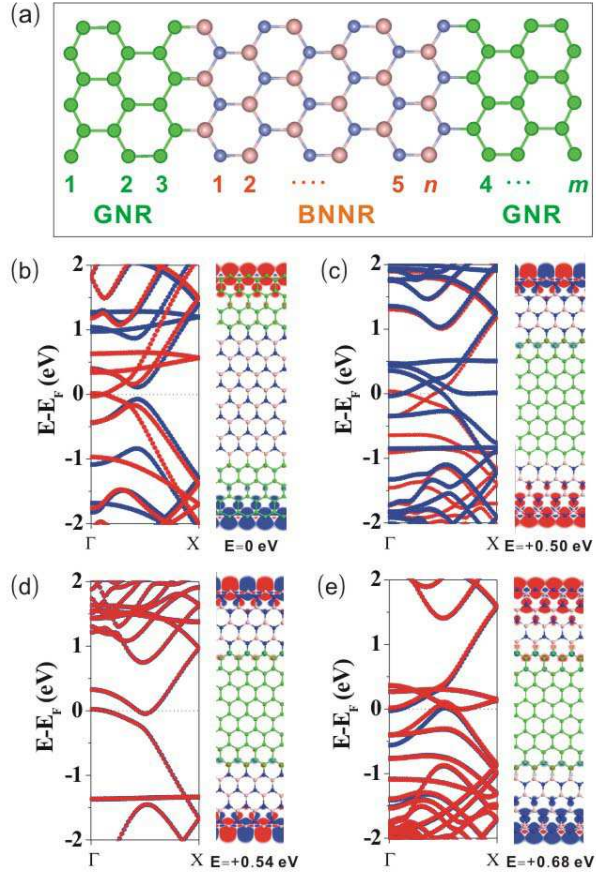}
\caption{(a) The schematic structure of (\emph{m}, \emph{n}) ZBNCNR
with \emph{m} and \emph{n} zigzag chains for the BNNR and GNR,
respectively. (b-e) The spin-polarized band structures of (b) (8, 8)
ZBNCNR-CC, (c) (8, 8) ZBNCNR-BN, (d) (8, 8) ZBNCNR-BB, and (e) (8,
8) ZBNCNR-NN. The corresponding optimized atomic structures
associated with the spatial distribution of (ground state) spin
densities and the relative (ground state) total energies are also
presented in (b)-(e). Blue and red colors represent the spin-up
and spin-down states, respectively. The Fermi level is set to zero.}
\end{figure}

We now turn to ZBNCNRs. According to our prediction, C prefers to
form the zigzag edges outside of BN. We take (8, 8) ZBNCNRs as
examples and different BNC configurations have been considered, as
shown in Figure 5. The case of C growing outside of BN is named as (8,
8) ZBNCNR-CC for short, as shown in Figure 5b. There are three
different configurations for the cases of BN growing outside of C:
(1) one edge is B-edge, and the other edge is N-edge (ZBNCNR-BN,
Figure 5c), (2) both edges are B-edges (ZBNCNR-BB, Figure 5d), (3) both
edges are N-edges (ZBNCNR-NN, Figure 5e). Because specific magnetic
orderings were found along the zigzag edges of GNRs and bare
BNNRs\cite{Son-2006,Zheng-2008,Barone-2008}, it is expected that
spin-polarization will also play an important role in ZBNCNRs. In
order to search for the most stable spin configurations, the total
energies of ZBNCNRs with different initial spin orderings are
calculated in double unit cell. The spin configurations that are
most energetically favorable are displayed in Figure 5b-5e: for
ZBNCNR-CC, spins have ferromagnetic ordering at each C-edge and
antiferromagnetic coupling between two C-edges (Figure 5b); for
ZBNCNR-BN, spins have antiferromagnetic ordering at the B-edge and
ferromagnetic ordering at the N-edge (Figure 5c); for ZBNCNR-BB, spins
have antiferromagnetic ordering at both B-edges (Figure 5d); for
ZBNCNR-NN, spins have ferromagnetic ordering at each N-edge and
antiferromagnetic coupling between two N-edges (Figure 5e). The total
energy of ZBNCNR-CC is more favorable than ZBNCNR-BN, ZBNCNR-BB, and
ZBNCNR-NN by 0.50, 0.54, and 0.68 eV per double unit cell (i.e.,
$\sim$ 0.050, 0.054 and 0.068 eV/\AA), respectively, consistent with
our prediction that C prefers to form the zigzag edges in BNCNRs.

\begin{figure}[tbp]
\includegraphics[width=8.0cm]{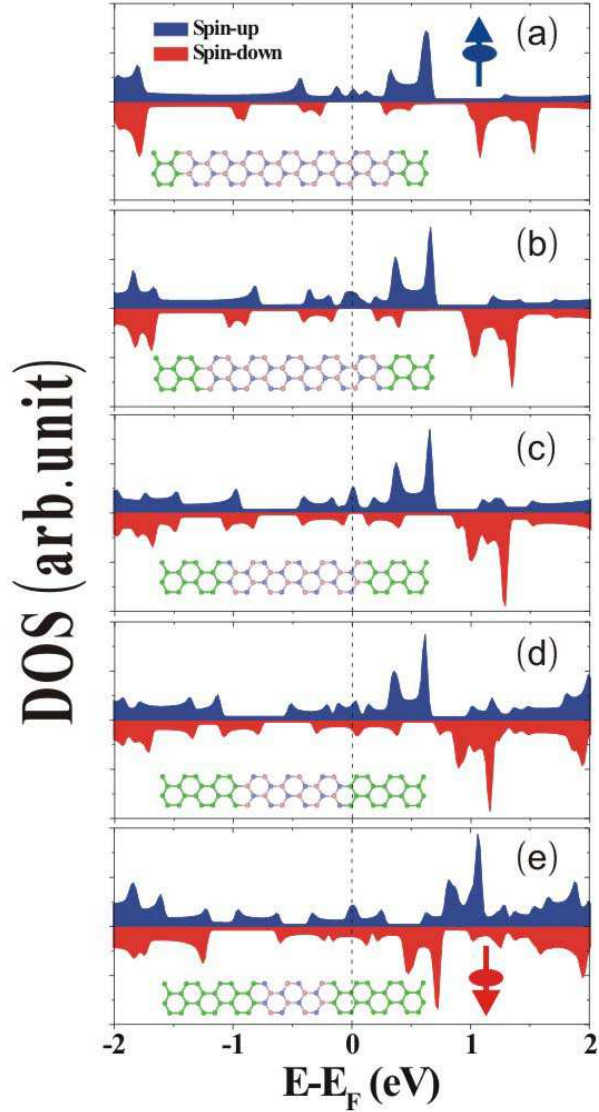}
\caption{The density of states of (a) (12, 4) ZBNCNR-CC, (b) (10, 6)
ZBNCNR-CC, (c) (8, 8) ZBNCNR-CC, (d) (6, 10) ZBNCNR-CC, and (e) (4,
12) ZBNCNR-CC. The insets show the corresponding atomic structures.
Blue and red colors represent spin-up and spin-down states,
respectively. The Fermi level is set to zero.}
\end{figure}

It is interesting to see that the ground state of (8, 8) ZBNCNR-CC
exhibits intrinsic half-metallic behavior (Figure 5b), with an
apparent gap ($\sim$ 0.2 eV) for the spin-up state and two bands
crossing the Fermi level for the spin-down state. This mainly
results from the large chemical potential difference between the C-B
and C-N boundaries as well as the hybridization of the orbitals of
C, B and N atoms at BN-C boundaries\cite{Bing-2010,
Kan-Li-Dutta-Pruneda}. Although there are already some reports on
half-metallicity and ferromagnetism in various BN\cite{Zheng-2008, Du-Zhang-Chen} and hybrid BNC structures\cite{Bing-2010, Kan-Li-Dutta-Pruneda}, the feasibility of realizing these structures in practice (i.e., the
stability of these structures) were unknown. Our work presents some
new insights on understanding these hybrid BNC structures. Different
from (8, 8) ZBNCNR-CC, (8, 8) ZBNCNR-BN (Figure 5c) is a ferromagnetic
metal with $\sim$ 2 $\mu$B net magnetic moment per double unit cell,
while (8, 8) ZBNCNR-BB (Figure 5d) and (8, 8) ZBNCNR-NN (Figure 5e) are
antiferromagnetic metal with zero net magnetic moment per double
unit cell. These results indicate that the electronic and magnetic
properties of ZBNCNRs are strongly dependent on the detailed
hybridized structures, which may be achieved in experiments under
specific growth conditions.

Moreover, the existence of half-metallicity as well as the value of
half-metallic gap in ZBNCNR-CC depends strongly on the C/BN ratio
(i.e., \emph{n}/\emph{m}), as shown in Figure 6. Clearly, the
half-metallic gap decreases from 0.27 eV (Figure 6a) to 0.19 eV (Figure
6b) to 0.12 eV (Figure 6c) by increasing the C/BN ratio from 33.3\%
to 60\% to 100\%, respectively. These half-metallic gaps are large
enough for room-temperature operation. ZBNCNR-CC will convert from
half-metal to ferromagnetic metal as the C/BN ratio is further
increased (larger than 100\%), as presented in Figure 6d and 6e. Thus,
the electronic and magnetic properties of ZBNCNRs
could be precisely modulated by changing the C/BN ratio.

\section{Summary}

In conclusion, using spin-polarized DFT calculations, we have
systemically studied the edge stability (edge energy and edge
stress) of single-layer BN. Our results demonstrate that the edges
of ABNNRs are more stable than those of ZBNNRs. ABNNRs are under
compressive edge stress but ZBNNRs are under tensile stress. The
intrinsic spin-polarization and edge adsorption of H could stabilize
the edges of BNNRs. Furthermore, the different edge stability
between BN and graphene may be used to provide some guiding
principles for designing the specific structures of hybrid BNC. For
examples, BN is expected to grow outside of C in ABNCNRs, while the
reverse is true in ZBNCNRs. The hybrid BNCNRs show rich electronic
and magnetic properties depending strongly on their detailed
structures and the atomic C and BN ratio. Our findings can be useful
for developing nanoscale electronics and spintronics.

\section*{Acknowledgments}

The work at Tsinghua was supported by the Ministry of Science and
Technology of China (Grant Nos. 2006CB605105 and 2006CB0L0601), and
the National Natural Science Foundation of China; The work at Utah
was supported by DOE.

\newpage


\end{document}